\documentclass[aps,prd,reprint]{revtex4-1}
\usepackage{blindtext}
\usepackage{graphicx}
\usepackage{color}

\usepackage{amsmath}
\usepackage{amssymb}

\newcommand{\simlt}
{\lower.5ex\hbox{\ltsima}}
\newcommand{\simgt}
{\lower.5ex\hbox{\gtsima}}

\begin{document}
\def\bigint{{\displaystyle\int}}
\def\simlt{\stackrel{<}{{}_\sim}}
\def\simgt{\stackrel{>}{{}_\sim}}

\title{Generalized Tsirelson's bound from parity symmetry considerations}
\author{David H. Oaknin}
\affiliation{Rafael Ltd, IL-31021 Haifa, Israel, \\
e-mail: { d1306av@gmail.com}}

\begin{abstract}
The Bell experiment is a random game with two binary outcomes whose statistical correlation is given by $E_0(\Theta)=-\cos(\Theta)$, where $\Theta \in [-\pi, \pi)$ is an angular input that parameterizes the game setting. The correlation function $E_0(\Theta)$ belongs to the affine space ${\cal H} \equiv \left\{E(\Theta)\right\}$ of all continuous and differentiable periodic functions $E(\Theta)$ that obey the parity symmetry constraints $E(-\Theta)=E(\Theta)$ and $E(\pi-\Theta)=-E(\Theta)$ with $E(0)=-1$ and, furthermore, are strictly monotonically increasing in the interval $[0, \pi)$. Here we show how to build explicitly local statistical models of hidden variables for random games with two binary outcomes whose correlation function $E(\Theta)$ belongs to the affine space ${\cal H}$. This family of games includes the Bell experiment as a particular case. Within this family of random games, the Bell inequality can be violated beyond the Tsirelson bound of $2\sqrt{2}$ up to the maximally allowed algebraic value of 4. In fact, we show that the amount of violation of the Bell inequality is a purely geometric feature. 
\end{abstract}


\maketitle

In the Bell experiment a source emits pairs of particles whose polarizations are arranged in an entangled state \cite{Bohm,EPR}:
\begin{equation}
\label{Bell_state}
| \Psi_{\Phi} \rangle = \frac{1}{\sqrt{2}} \left(|\uparrow \rangle^{(A)} \ |\downarrow\rangle^{(B)} - e^{i \Phi} \ |\downarrow \rangle^{(A)} \ |\uparrow\rangle^{(B)}\right), 
\end{equation}
where $\left\{|\uparrow\rangle, \ |\downarrow\rangle\right\}^{(A,B)}$ are bases of single-particle eigenstates of Pauli operators $\sigma_Z^{(A,B)}$ along locally defined Z-axes for each one of them. The two emitted particles travel off the source in opposite directions towards two widely separated detectors, which test their polarizations. The orientation of each one of the detectors can be freely and independently set along any arbitrary direction in the XY-plane perpendicular to the locally defined Z-axis.  Upon detection each particle causes a binary response of its detector, either $+1$ or $-1$. Thus, each detected pair of entangled particles produces an outcome in the space of possible events $\left\{(-1,-1), (-1,+1), (+1,-1), (+1,+1)\right\}$. In general, the statistical correlation  between the outcomes of the two detectors in a long sequence of repetitions of the Bell experiment is a number $E_0(\Theta)$ in the interval $\left[-1,+1\right]$, which depends on the relative angle $\Theta$ between the orientations of the two detectors. By definition, the correlation function $E_0(\Theta)$ is thus a periodic function of the relative angle $\Theta$ with a period of $2\pi$. Quantum mechanics predicts that the correlation is given by:
\begin{equation}
\label{correlation}
E_0(\Theta) \equiv -\cos(\Theta),
\end{equation}  
where the relative angle is defined with respect to a reference setting of the detectors at which $\Theta=0$ and the outcomes of the two detectors are fully anticorrelated, $E_0(\Theta=0)=-1$.

It can be readily shown that the following inequality holds for any set of values $\left(\Theta_1, \Theta_2, \delta\right) \in \left(\mathbb{R} \ \mbox{mod} \left[-\pi, \pi\right)\right)^3$ \cite{Tsirelson}: 
\begin{eqnarray}
\label{Tsirelson}
\left|{\cal F}_{E_0(\Theta)}(\Theta_1, \Theta_2, \delta)\right| \le 2 \sqrt{2},  
\end{eqnarray}   
where
\begin{equation}
\nonumber
{\cal F}_{E(\Theta)}(\Theta_1, \Theta_2, \delta) \equiv E(\Theta_1) + E(\Theta_2) +  E(\Theta_1-\delta) - E(\Theta_2-\delta).
\end{equation}
Moreover, the bound (\ref{Tsirelson}) - known as the Tsirelson bound - is saturated for certain values of the parameters. For example, $|{\cal F}_{E_0(\Theta)}(+\frac{\pi}{4},-\frac{\pi}{4},+\frac{\pi}{2})| = \frac{1}{\sqrt{2}}
+\frac{1}{\sqrt{2}}+\frac{1}{\sqrt{2}}-(-\frac{1}{\sqrt{2}})=2\sqrt{2}$. This bound is thought to be a characteristic feature of the quantum theory \cite{PopescuRorhlich} associated to some fundamental physical principle, {\it e.g.} the so-called information causality principle \cite{Pawlowski}, which would explain why quantum correlations go only up to the Tsirelson bound, $2\sqrt{2}$, and not beyond it to the algebraically possible maximum value of $4$. In this paper we explore this question from a purely geometric perspective.

First, we shall show that any function $E(\Theta)$ that fulfills the symmetry constraints:
\begin{eqnarray}
\label{constraint1}
& & E(\Theta=0) = -1, \\  
\label{constraint2}
& & E(-\Theta) = E(\Theta), \\
\label{constraint3}
& & E(\pm \pi \mp \Theta) =-E(\Theta),
\end{eqnarray}
and it is everywhere continuous and, at least, twice-differentiable and, moreover,
\begin{eqnarray}
\label{constraint0}
& & E''(\Theta) > 0, \ \  \forall \ \Theta \in [0, \pi/2],
\end{eqnarray}
fulfills the constraint
\begin{equation}
\label{Generalized}
\left|{\cal F}_E(\Theta_1, \Theta_2, \delta)\right| \le \left|{\cal F}_E(+\frac{\pi}{4},-\frac{\pi}{4},+\frac{\pi}{2})\right| = 4 \left| E(\pi/4)\right|.
\end{equation}

\begin{figure}
\vspace{-1.0in}
\begin{centering}
\includegraphics[height=12cm]{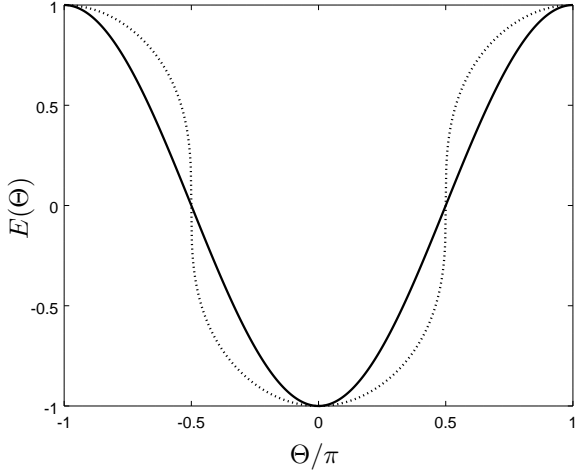}
\end{centering}
\fontsize{10}{0}
\selectfont\put(-129.,84.){\makebox(0,0)[t]{\textcolor[rgb]{0,0,0}{{\bf $\Theta/\pi$}}}}
\selectfont\put(-235.,173.){\rotatebox{90}{\makebox(0,0)[b]{\textcolor[rgb]{0,0,0}{{\bf $E(\Theta)$}}}}}
\vspace{-1.1in}
\caption{Two examples of possible correlation functions for the random games with two binary outcomes described by eq. (\ref{constraint1}-\ref{constraint0}): a) $E_0(\Theta)=-\cos(\Theta)$ (solid line) and b) $E_1(\Theta)=-\mbox{sign}(\cos(\Theta)) \cdot\left|\cos(\Theta)\right|^{1/3}$ (dotted line). In the first example $4\left|E_0(\pi/4)\right|=2 \sqrt{2}$, while in the second $4\left|E_0(\pi/4)\right|=2^{11/6}$.}
\end{figure}

Thus, we see that the amount of violation of the Bell inequality is directly related to the statistical correlation between the outcomes of the two detectors when they are oriented at a relative angle $\Theta=\pi/4$. In particular, for the correlation function (\ref{correlation}) we have $E_0(\pi/4)=-\cos(\pi/4)=-1/\sqrt{2}$ and, hence, from (\ref{Generalized}) we obtain the Tsirelson bound (\ref{Tsirelson}).

In order to prove the generalized constraint (\ref{Generalized}) we notice that ${\cal F}_{E(\Theta)}(\Theta_1, \Theta_2, \delta)$ is a periodic, continuous and differentiable function over its three dimensional domain, and, therefore, it is bounded from above and below and, furthermore, it reaches its extrema. At this extremal points the following conditions must be fulfilled:
\begin{eqnarray}
\begin{array}{cccccccccc}
\nonumber
\frac{\partial F_{E(\Theta)}(\Theta_1, \Theta_2, \delta)}{\partial \Theta_1}&=&E'(\Theta_1) & + & E'(\Theta_1 - \delta) & = & 0, \\
\nonumber
\frac{\partial F_{E(\Theta)}(\Theta_1, \Theta_2, \delta)}{\partial \Theta_2}&=&E'(\Theta_2) & - & E'(\Theta_2 - \delta) & = & 0, \\
\nonumber
\frac{\partial F_{E(\Theta)}(\Theta_1, \Theta_2, \delta)}{\partial \delta}&=&E'(\Theta_2 - \delta) & - & E'(\Theta_1 - \delta) & = & 0,
\end{array}
\end{eqnarray}
which imply that at the extremal points 
\begin{eqnarray}
- E'(\Theta_1) = E'(\Theta_2) = E'(\Theta_1 - \delta) = E'(\Theta_2 - \delta).
\end{eqnarray}
Hence, as long as $\left|E(\pi/4)\right| > 0.5$ the absolute extrema are located at
\begin{equation}
\Theta_1=\pm \frac{\pi}{4}, \ \ \Theta_2=\mp \frac{\pi}{4}, \ \ \delta=\pm \frac{\pi}{2},
\end{equation}
and constraint (\ref{Generalized}) readily follows. 

Thus, the Tsirelson bound is, in fact, a bound on the maximal correlation between the outcomes of the two detectors when they are set at a relative angle $\Theta=\pi/4$. Notwithstanding, since constraints (\ref{constraint1}-\ref{constraint0}) also imply 
\begin{eqnarray*}
\left|E(\Theta)\right| > \left|E(\pi/4)\right|, \ \ \  |\Theta| < \pi/4,  
\end{eqnarray*}
it seems difficult to accept that some fundamental physical principle bounds the correlation at the relative angle $\Theta=\pi/4$ while the same bound may be violated at smaller relative angles. 

In fact, we shall now show how to build an explicitly local statistical model of hidden variables for any random game with two binary outcomes whose correlation is described by a correlation function $E(\Theta)$ within the affine space ${\cal H}$ defined by the constraints (\ref{constraint1}-\ref{constraint0}). This family of random games show that the Bell inequality may be violated beyond the Tsirelson bound up to the maximally allowed algebraic value without requiring any violation of locality.

First, we notice that constraints (\ref{constraint1}-\ref{constraint0}) imply that the correlation function $E(\Theta)$ is continuous and strictly monotonically increasing in the interval $[0, \pi]$, with $E(0)=-1$ and $E(\pi)=+1$ and, hence, there exists a unique function $\chi(E)$ defined in the interval $[-1,+1]$, such that 

\begin{equation}
\chi(E(\Theta))=\Theta, \ \ \ \ \forall \Theta \in [0, \pi]. 
\end{equation}
Furthermore, the function $\chi(E)$ is continuous and differentiable over its domain of definition $[-1,+1]$. 

We now consider a continuous infinite set of possible hidden configurations distributed over a unit circle ${\cal S}$. Each one of the two detectors involved in the random game defines a set of coordinates over the unit circle, which we shall label, respectively, as $\lambda_A \in [-\pi, +\pi)$ and $\lambda_B \in [-\pi, +\pi)$. Since the two sets of coordinates parameterize the same space of possible configurations there must exist a transformation law that relates them:

\begin{equation}
\lambda_B = - L(\lambda_A; \Theta),
\end{equation}
which may depend parameterically on the relative angle $\Theta$ between the orientations of the two detectors, measured with respect to the reference setting $\Theta=0$ at which their outcomes are fully anti-correlated (that is, $E(\Theta=0)=-1$). We define the transformation law as follows:

\begin{itemize}
\item If  $\Theta \in [0, \pi)$, 
\begin{eqnarray}
\label{Oaknin_transformation}
\hspace{-0.15in}
L(\lambda; \Theta) =  
\left\{
\begin{array}{c}
\hspace{-0.2in} q(\lambda-\Theta) \cdot \chi\left(-E(\Theta) - E(\lambda) + 1 \right), \\ \hspace{0.88in} \mbox{if}  \hspace{0.1in} -\pi \hspace{0.13in} \le  \lambda < \Theta-\pi, \\
\hspace{-0.32in} q(\lambda-\Theta) \cdot \chi\left(E(\Theta) + E(\lambda) + 1 \right), \\ \hspace{0.685in} \mbox{if}  \hspace{0.02in} \Theta-\pi \hspace{0.08in} \le \lambda < \hspace{0.105in} 0, \\
\hspace{-0.32in} q(\lambda-\Theta) \cdot \chi\left(E(\Theta) - E(\lambda) - 1 \right), \\ \hspace{0.69in} \mbox{if}  \hspace{0.25in} 0 \hspace{0.15in} \le \lambda < \ \Theta, \\
\hspace{-0.2in} q(\lambda-\Theta) \cdot \chi\left(-E(\Theta) + E(\lambda) - 1 \right), \\ \hspace{0.72in} \mbox{if}  \hspace{0.21in} \Theta  \hspace{0.16in} \le  \lambda  < +\pi, \\
\end{array}
\right.
\end{eqnarray}
\item If  $\Theta \in [-\pi, 0)$, 
\begin{eqnarray}
\label{Oaknin_transformation_Inv}
\hspace{-0.15in}
L(\lambda; \Theta) =  
\left\{
\begin{array}{c}
\hspace{-0.2in} q(\lambda-\Theta) \cdot \chi\left(-E(\Theta) + E(\lambda) + 1 \right), \\ \hspace{0.650in} \mbox{if}   \hspace{0.11in} -\pi \hspace{0.15in} \le \lambda < \Theta, \\
\hspace{-0.32in} q(\lambda-\Theta) \cdot \chi\left(E(\Theta) - E(\lambda) + 1 \right), \\ \hspace{0.66in} \mbox{if}   \hspace{0.30in} \Theta \hspace{0.1in} \le \lambda < \hspace{0.05in} 0, \\
\hspace{-0.32in} q(\lambda-\Theta) \cdot \chi\left(E(\Theta) + E(\lambda) - 1 \right), \\ \hspace{0.92in} \mbox{if}  \hspace{0.23in} 0 \hspace{0.22in} \le \lambda < \Theta +\pi, \\
\hspace{-0.2in} q(\lambda-\Theta) \cdot \chi\left(-E(\Theta) - E(\lambda) - 1 \right), \\ \hspace{0.75in} \mbox{if}  \hspace{0.16in} \Theta +\pi \hspace{0.00in} \le \lambda < +\pi, \\
\end{array}
\right.
\end{eqnarray}
\end{itemize}
with 
\begin{eqnarray*}
q(\lambda-\Theta) = \mbox{sign}((\lambda - \Theta) \mbox{mod} ([-\pi, \pi))).
\end{eqnarray*}
In Fig. 1 two different possible correlation functions are plotted against each other: $E_0(\Theta)=-\cos(\Theta)$ and $E_1(\Theta)=-\mbox{sign}(\cos(\Theta))\cdot\cos^{1/3}(\Theta)$. In Fig. 2 the transformation $\lambda_B=-L(\lambda_A; \Theta)$ is graphically shown for the particular case $\Theta = \pi/3$ when: a) $E(\Theta)=E_0(\Theta)=-\cos(\Theta)$ and b) $E(\Theta)=E_1(\Theta)=-\mbox{sign}(\cos(\Theta)) \cdot\left|\cos(\Theta)\right|^{1/3}$ .  

\begin{figure}
\vspace{-1.0in}
\begin{centering}
\includegraphics[height=12cm]{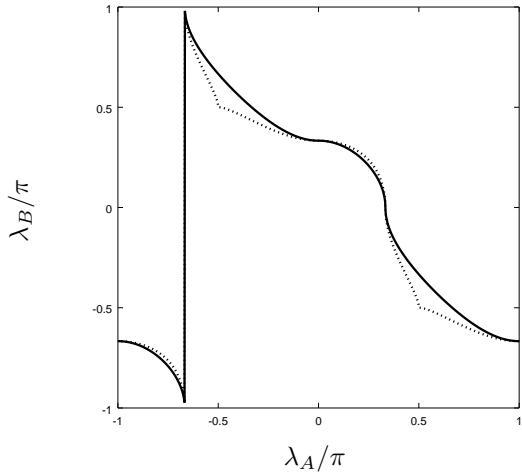}
\end{centering}
\fontsize{10}{0}
\selectfont\put(-129.,84.){\makebox(0,0)[t]{\textcolor[rgb]{0,0,0}{{\bf $\lambda_A/\pi$}}}}
\selectfont\put(-235.,173.){\rotatebox{90}{\makebox(0,0)[b]{\textcolor[rgb]{0,0,0}{{\bf $\lambda_B/\pi$}}}}}
\vspace{-1.1in}
\caption{Transformation law $\lambda_B = -L(\lambda_A; \Delta)$ with $\Delta=\pi/3$ for two examples of possible correlation functions: $E_0(\Theta)=-\cos(\Theta)$ (solid line) and $E_1(\Theta)=-\mbox{sign}(\cos(\Theta)) \cdot\left|\cos(\Theta)\right|^{1/3}$ (dotted line).}
\end{figure}

It is straightforward to check that the transformation law (\ref{Oaknin_transformation},\ref{Oaknin_transformation_Inv}) fulfills the differential relationship
\begin{equation}
\label{free}
\left|dE(\lambda_B)\right|  = \left|dE(\lambda_A)\right|,
\end{equation}
and, hence,
\begin{equation}
\label{free}
d\lambda_B \cdot \left| E'(\lambda_B)\right| = d\lambda_A \cdot  \left|E'(\lambda_A)\right|.
\end{equation}
Therefore, if we define the (density of ) probability of each configuration to happen in every single realization as
\begin{equation}
\rho(\lambda) = \frac{1}{4} \left|E'(\lambda)\right|,
\end{equation}  
so that 
\begin{equation}
\int_{-\pi}^{\pi} d\lambda \ \rho(\lambda) = 2 \int_{0}^{\pi} d\lambda \ \rho(\lambda) = \frac{1}{2} \int_{0}^{\pi} d\lambda E'(\lambda) = 1
\end{equation}
and we have from eq. (\ref{free}) that 
\begin{equation}
\label{free2}
d\lambda_B \cdot \rho(\lambda_B) = d\lambda_A \cdot  \rho(\lambda_A),
\end{equation}
which states that, as required, the probability does not change under a coordinate transfromation, so that 'free-will' is fulfilled.

Finally, we define the response function of the detectors as

\begin{equation}
s_A = \tau(\lambda_A), \hspace{0.4in} s_B = \tau(\lambda_B),
\end{equation}
with
\begin{eqnarray}
\tau(\lambda) = \left\{
\begin{array}{cccccc}
+1, & \mbox{if} & \lambda \in [0, +\pi), \\
-1, & \mbox{if} & \lambda \in [-\pi, 0).
\end{array}
\right.
\end{eqnarray}
Thus, the binary outcomes of the two detectors define a partition of the phase space ${\cal S}$ of all the possible hidden configurations into four coarse subsets,
\begin{eqnarray*}
\label{four_subsets}
(s_A=+1; s_B=+1) & \Longleftrightarrow & \lambda_A \in [0, \Theta) \\
(s_A=+1; s_B=-1) & \Longleftrightarrow & \lambda_A \in [\Theta, \pi) \\
(s_A=-1; s_B=+1) & \Longleftrightarrow & \lambda_A \in [\Theta-\pi, 0) \\
(s_A=-1; s_B=-1) & \Longleftrightarrow & \lambda_A \in [-\pi, \Theta-\pi),
\end{eqnarray*}
where we have assumed without any loss of generality that $\Theta \in [0, \pi)$. Each one of these four coarse subsets happen with a probability given by: 
\begin{eqnarray*}
\begin{array}{cccccc}
p\left(+1,+1\right) & =  & \int_0^{\Theta} \rho(\lambda_A) \ d\lambda_A \hspace{0.15in} & = \ \frac{1}{4}\left(1 + E(\Theta)\right), \vspace{0.1in} \\ 
p\left(+1,-1\right)  & =  & \int_{\Theta}^{\pi} \rho(\lambda_A) \ d\lambda_A \hspace{0.12in} & = \ \frac{1}{4}\left(1 - E(\Theta)\right), \vspace{0.1in} \\
p\left(-1, +1\right) & = & \int_{\Theta-\pi}^{0} \rho(\lambda_A) \ d\lambda_A & = \ \frac{1}{4}\left(1 - E(\Theta)\right), \vspace{0.1in} \\
p\left(-1,-1\right) & = & \int_{-\pi}^{\Theta-\pi} \rho(\lambda_A) \ d\lambda_A & = \ \frac{1}{4}\left(1 + E(\Theta)\right).
\end{array}
\end{eqnarray*}
It is then straightforward to notice that the model reproduces the desired correlation function:

\begin{eqnarray}
\nonumber
 p\left(+1,+1\right) - p\left(+1,-1\right) - p\left(-1,+1\right) + p\left(-1,-1\right) = \\
\nonumber
= E(\Theta). \hspace{0.0in}
\end{eqnarray}
\

The random games with two binary outcomes discussed in this paper, of which the Bell game is a particular case, are explictly local in the most strict sense of the term. Nonetheless, within this family the amount of violation of the Bell inequality is not constrained by the Tsirelson bound and can indeed reach the maximally allowed algebraic value. In fact, we have shown that this amount is a purely geometric feature.  The analysis presented in this paper is a follow-up of ideas previously discussed in \cite{oaknin1,oaknin2}.


\begin{references}


\bibitem{Bohm} D.~Bohm,  Quantum Theory {\bf 1951}, Prentice-Hall, New York.

\bibitem{EPR} A.~Einstein, B.~Podolsky and N.~Rosen, "Can the quantum mechanical description of physical reality be considered complete ?",  Phys. Rev. {\bf 1935}, 47, 777-780, DOI: 10.1103/Phys.Rev.47.777.

\bibitem{Tsirelson} B. S.~Cirelson, "Quantum generalizations of Bell's inequality", Lett. Math. Phys. {\bf 1980}, 4, 93.

\bibitem{PopescuRorhlich} S.~Popescu and D.~Rohrlich, "Quantum nonlocality as an axiom", Foundations of
Physics, {\bf 1994}, 24, 379–385.

\bibitem{Pawlowski} M.~Pawłowski, T.~Paterek, D.~Kaszlikowski, V.~Scarani, A.~Winter, M.~Zukowski,  ˙
"Information causality as a physical principle", Nature, {\bf  2009}, 461, 1101–1104.

\bibitem{oaknin1} D.H.~Oaknin, "Solving the EPR paradox: an explicit statistical local model of hidden variables for the singlet state", arXiv:1411.5704.

\bibitem{oaknin2} D.H.~Oaknin, "Solving the EPR paradox: geometric phases in gauge theories", Frontiers in Physics 8:142 (2020), arXiv:1912.06349.

\end{references}
\end{document}